\documentclass[twocolumn, trackchanges]{aastex62}

\usepackage{multirow}
\usepackage{bm}
\usepackage{amsmath}
\usepackage{graphicx}

\newcommand{\myemail}{m.cornachione@utah.edu}
\newcommand{\Romannumeral}[1]{\uppercase\expandafter{\romannumeral #1\relax}}

\shorttitle{Spectro-Perfectionism for Radial Velocities}
\shortauthors{Cornachione et al.}

\begin{document}

\title{A Full Implementation of Spectro-Perfectionism for Precise Radial Velocity Exoplanet Detection: A Test Case With the MINERVA Reduction Pipeline\footnote{This is an author-created, un-copyedited version of an article accepted for publication in Publications of the Astronomical Society of the Pacific. The publisher is not responsible for any errors or omissions in this version of the manuscript or any version derived from it. The Version of Record is available online at https://dx.doi.org/10.1088/1538-3873/ab4103.}}

\author[0000-0003-1012-4771]{Matthew A. Cornachione}
\affiliation{Department of Physics and Astronomy, University of Utah, 115 South 1400 East, Salt Lake City, UT 84112, USA}
\affiliation{Department of Physics, United States Naval Academy, 572C Holloway Rd., Annapolis, MD 21402, USA}

\author[0000-0002-9836-603X]{Adam S. Bolton}
\affiliation{National Optical Astronomy Observatory, 950 N. Cherry Ave., Tucson, AZ 85719, USA}
    
\author[0000-0003-3773-5142]{Jason D.\ Eastman}
\affiliation{Center for Astrophysics \textbar \ Harvard \& Smithsonian, 60 Garden St, Cambridge, MA 02138, USA}
    
\author[0000-0003-1928-0578]{Maurice L. Wilson} 
\affiliation{Center for Astrophysics \textbar \ Harvard \& Smithsonian, 60 Garden St, Cambridge, MA 02138, USA}

\author[0000-0002-6937-9034]{Sharon X. Wang}
\affiliation{Department of Terrestrial Magnetism, Carnegie Institution for Science, 5241 Broad Branch Road, NW, Washington, DC 20015, USA}

\author[0000-0001-9397-4768]{Samson A. Johnson} 
\affiliation{Department of Astronomy, The Ohio State University, 140 West 18th Avenue, Columbus, OH 43210, USA}

\author{David H. Sliski} 
\affiliation{The University of Pennsylvania, Department of Physics and Astronomy, Philadelphia, PA, 19104, USA}

\author[0000-0002-8041-1832]{Nate McCrady}
\affiliation{Department of Physics and Astronomy, University of Montana, 32 Campus Drive, No. 1080, Missoula, MT 59812 USA}

\author[0000-0001-6160-5888]{Jason T. Wright}
\affiliation{Department of Astronomy and Astrophysics and Center for Exoplanets and Habitable Worlds, The Pennsylvania State University, University Park, PA 16802}

\author[0000-0002-8864-1667]{Peter Plavchan}
\affiliation{Department of Physics and Astronomy, George Mason University, 4400 University Drive, MSN 3F3, Fairfax, VA 22030, USA}

\author[0000-0002-1704-6289]{John Asher Johnson}
\affiliation{Center for Astrophysics \textbar \ Harvard \& Smithsonian, 60 Garden St, Cambridge, MA 02138, USA}

\author{Jonathan Horner}
\affiliation{Centre for Astrophysics, University of Southern Queensland, USQ Toowoomba, QLD 4350, Australia}

\author[0000-0001-9957-9304]{Robert A. Wittenmyer}
\affiliation{Centre for Astrophysics, University of Southern Queensland, USQ Toowoomba, QLD 4350, Australia}

    
\email{\myemail}

\begin{abstract}
We present a computationally tractable implementation of spectro-perfectionism, a method which minimizes error imparted by spectral extraction. We develop our method in conjunction with a full raw reduction pipeline for the MINiature Exoplanet Radial Velocity Array (MINERVA), capable of performing both optimal extraction and spectro-perfectionism.  Although spectro-perfectionism remains computationally expensive, our implementation can extract a MINERVA exposure in approximately $30\,\text{min}$.  We describe our localized extraction procedure and our approach to point spread function fitting.  We compare the performance of both extraction methods on a set of 119 exposures on HD122064, an RV standard star. Both the optimal extraction and spectro-perfectionism pipelines achieve nearly identical RV precision under a six-exposure chronological binning.
We discuss the importance of reliable calibration data for point spread function fitting and the potential of spectro-perfectionism for future precise radial velocity exoplanet studies.
\end{abstract}


\section{Introduction}




The radial velocity method has proven to be a powerful method for detecting and characterizing exoplanets \citep[e.g.,][]{camp1988a, lath1989a, mayo1995a, bara1996a, mayo2003a}. The next generation of surveys now strives to achieve a radial velocity precision of $10\text{--}30\,\text{cm/s}$ to enable the detection an Earth-mass planet in the habitable zone of a Sun-like star \citep{fisc2014a, fisc2016a}.  However, reaching such an extreme radial velocity precision presents immense challenges.  Many effects that have been negligible for previous studies cause apparent radial velocity shifts on the order of tens of cm/s, the same level as the target planets \citep{fisc2016a, jurg2016a, schw2016a}.  We focus here in particular on addressing software-based error introduced during spectral extraction. 

Optimal extraction \citep{horn1986a} has been the standard for radial velocity pipelines in recent years \citep[e.g.,][]{pisk2002a, angl2012a, chak2014a, ritt2014a, zech2014a, bern2015a, sosn2015a, roy2016a, brah2017a, zech2018a, bede2019a, bech2019a, ma2019a}.  While the performance of optimal extraction has been excellent for previous applications, it is known to be imperfect.  Optimal extraction relies on the assumption of the separability of the incident intensity into horizontal and vertical components, which is nearly, but never precisely, satisfied in real instruments \citep[][e.g.]{guan2015a}.  It also tends to degrade the resolution of Echelle spectra due to the tilt of orders projected on the detector. The influence of these effects is difficult to quantify, and depends on the specifics of each instrument, but they are expected to affect radial velocity precision at a level of $\sim10\,\text{cm/s}$.


A mathematically ``perfect'' extraction algorithm, which accounts for the true two-dimensional intensity profile, was developed by \citet{bolt2010a}.  This method, known as spectro-perfectionism (SP), was originally conceived for the extraction of faint-object spectra, but addresses the sources of imprecision inherent in optimal extraction.  Hence spectro-perfectionism may prove to be a crucial ingredient in extremely precise radial velocity (EPRV) survey pipelines.  

Unfortunately, as originally posed in \citet{bolt2010a}, spectro-perfectionism is computationally intractable.  We must therefore address this limitation before the method can be used in practice.  To this end, we explore the real-world capabilities of spectro-perfectionism in the context of the MINiature Exoplanet Radial Velocity Array (MINERVA) \citep[][accepted]{swif2015a, wils2019a}.  MINERVA is a dedicated exoplanet observatory which observes stars at a very high cadence.  The MINERVA project targets a precision of $0.8\,\text{m/s}$, with a currently achieved on-sky precision of $1.8\,\text{m/s}$ \citep{wils2019a}, where the performance gains of spectro-perfectionism start to become relevant.  This array therefore provides a perfect test for developing a practical spectro-perfectionism algorithm.

We lay out this paper as follows.  Section~\ref{sec:data} describes the MINERVA array and the KiwiSpec spectrograph as well as the science and calibration data.  Section~\ref{sec:pipeline} describes our calibration procedures and optimal extraction algorithm for the production pipeline.  In section~\ref{sec:sp}, we explain our techniques to implement spectro-perfectionism and in section~\ref{sec:results} we show the relative performance between the two extraction methods.
We discuss these findings in section~\ref{sec:discussion} and conclude with take-away points to address when considering SP pipelines for other instruments.

\section{MINERVA Instrument and Data}
\label{sec:data}

\begin{figure*}
    \centering
    \includegraphics[width=0.95\textwidth]{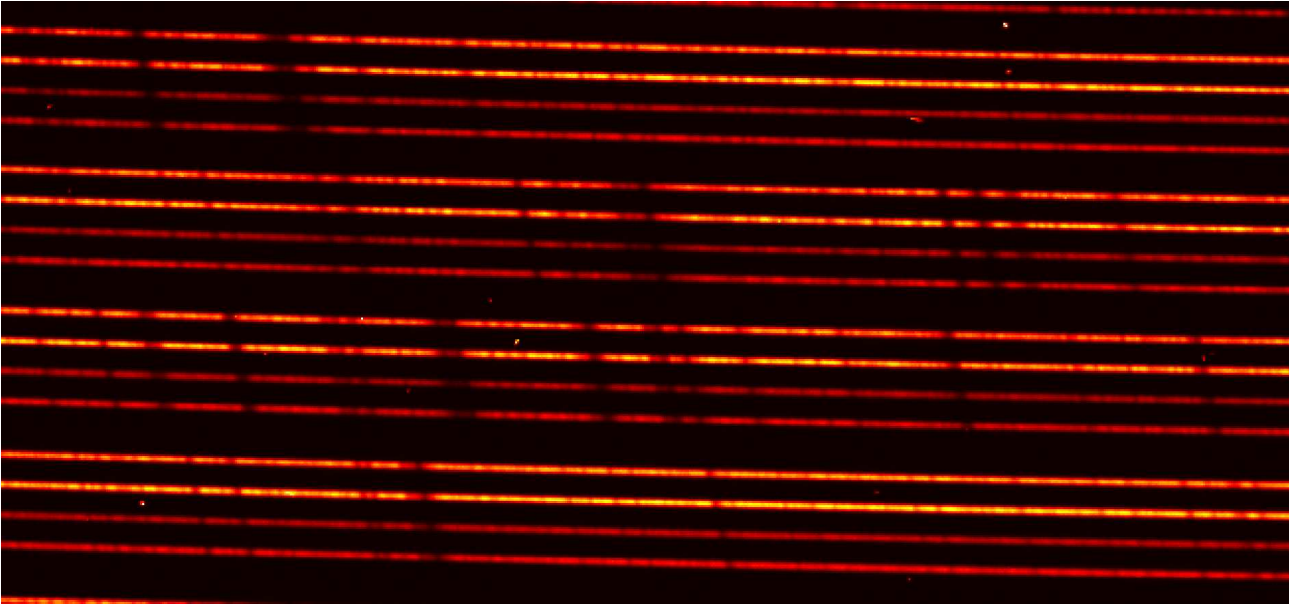}
    \caption{Zoom in on a section of the MINERVA CCD from a typical exposure.  Each bundle of four fibers observes the same region of the target spectrum, as is apparent from common absorption features.}
    \label{fig:minerva_ccd}
\end{figure*}

The MINERVA project is an array of four $0.7\,\text{m}$ telescopes located at Mt. Hopkins, Arizona \citep{swif2015a}. 
Each telescope can independently operate in photometric or spectroscopic mode.  For the purposes of this paper, we focus on the spectroscopic operation in which light is passed through four circular fibers to the KiwiSpec spectrograph \citep{swif2015a}.  This Echelle spectrograph has a nominal resolution of $R=80,000$ and is held at cryogenic temperatures and pressure-controlled for long-term stability of the instrument point spread function \citep{wils2019a}


Data are collected on a $2\text{k} \times 2\text{k}$ charge-coupled device (CCD) with  
a nominal gain of $1.3$ and readout noise of $3.63$.
The detector records 27 complete Echelle orders for each telescope, covering the wavelength range $4900\text{--}6460\text{\AA}$.  We show a portion of a typical raw science exposure in Figure~\ref{fig:minerva_ccd}.  An in-line iodine absorption cell is used for precision wavelength calibration, with the potential to achieve a precision of $0.8\,\text{m/s}$ \citep{wils2019a}.


We take nightly bias, dark, and slit flat frames.  For each, we collect and median stack eleven exposures to generate high signal-to-noise calibration images.  We also collect fiber flats and thorium-argon (ThAr) arc frames for MINERVA, but these require opening the spectrograph and so are only taken during major instrument reconfigurations, years apart.  Science exposures are taken throughout the night with an average of two exposures per target \citep{wils2019a}.  We also collect daytime sky exposures and check fiber guiding with a fiber acquisition unit (FAU).

\section{Raw Reduction Pipeline}
\label{sec:pipeline}

\begin{figure*}
    \centering
    \includegraphics[width=\textwidth]{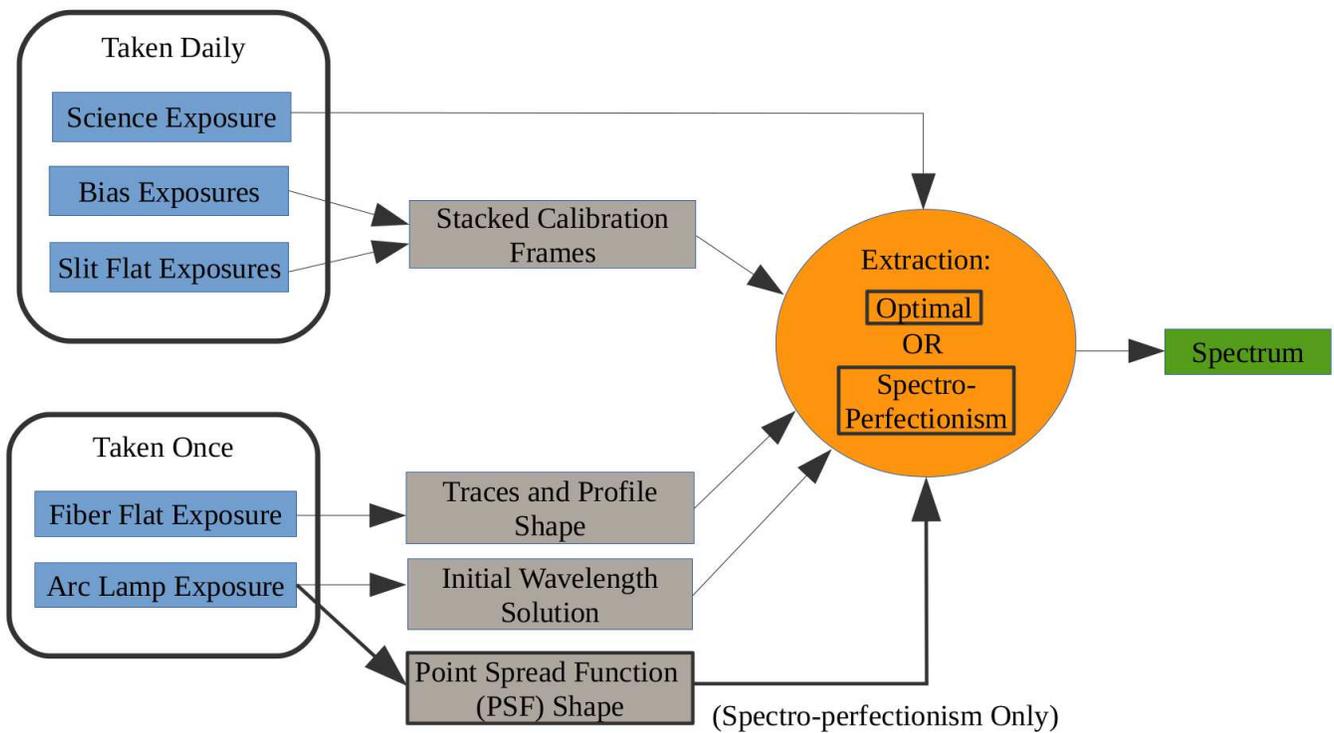}
    \caption{Schematic of the MINERVA raw reduction pipeline.  Both optimal extraction and spectro-perfectionism are illustrated.  MINERVA software is publicly available at https://github.com/MinervaCollaboration/minerva-pipeline.}
    \label{fig:pipeline}
\end{figure*}

\begin{figure}
    \centering
    \includegraphics[width=0.5\textwidth]{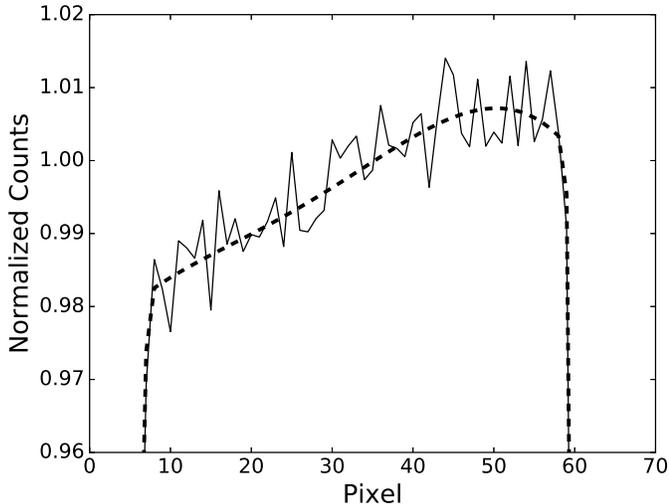}
    \caption{Cross-section of a typical MINERVA slit flat.  The raw stacked counts are shown with the solid line and the interpolated profile with the dotted line.}
    \label{fig:slit_flat}
\end{figure}

\begin{figure*}[t]
\gridline{\fig{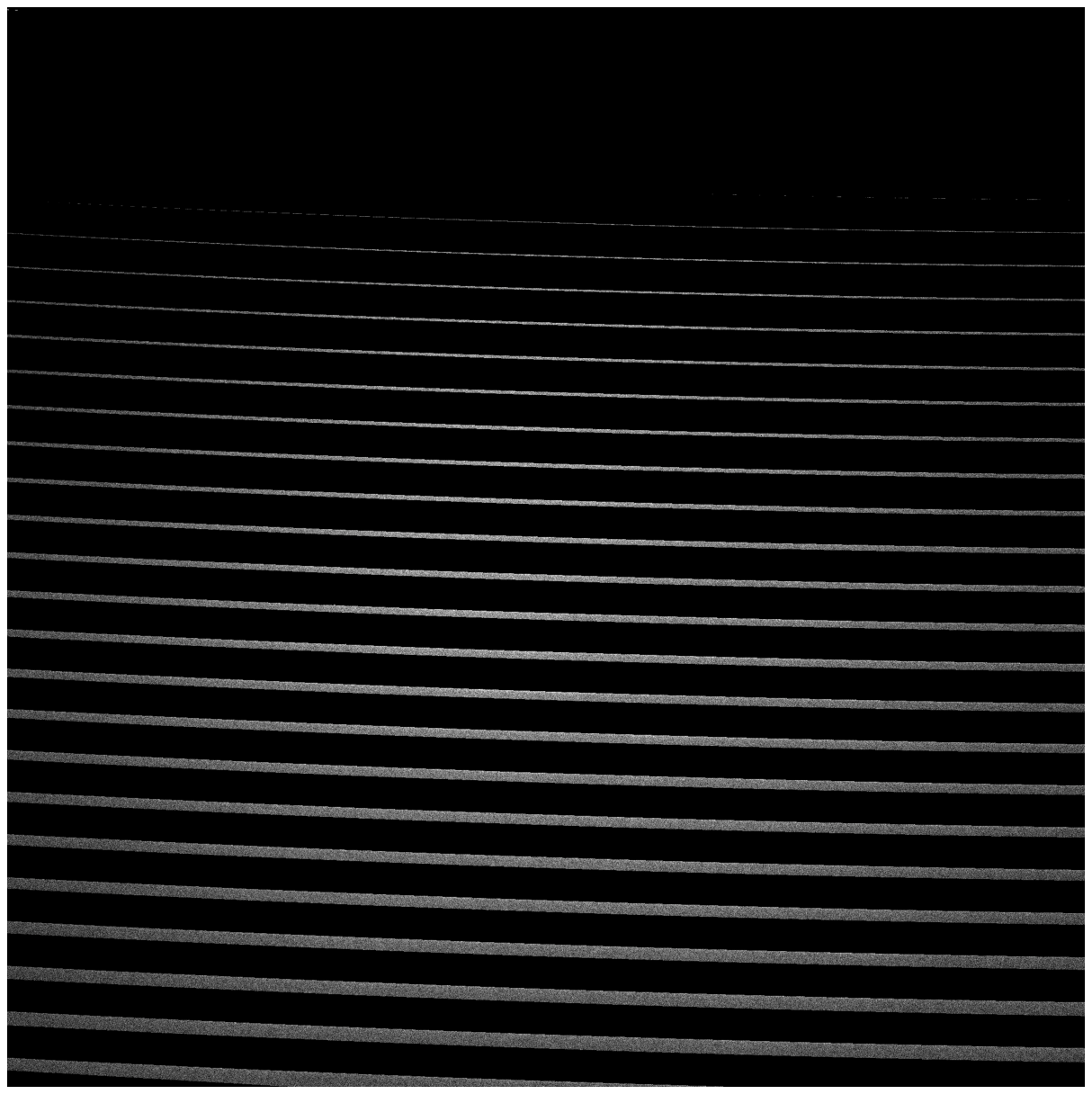}{0.5\textwidth}{(a)}
          \fig{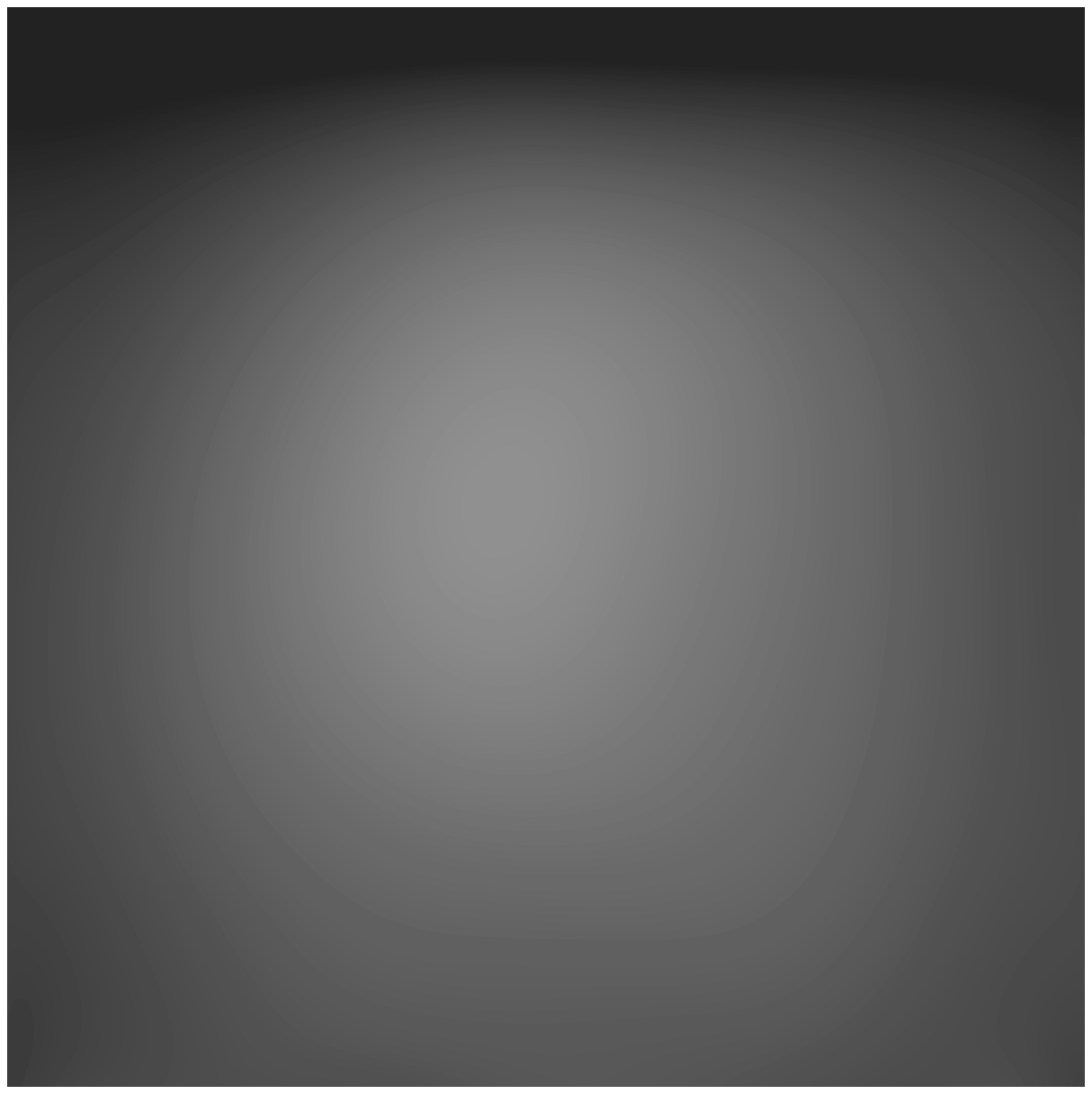}{0.5\textwidth}{(b)}}
  \caption{(a) Science exposure with fiber bundles masked, leaving only scattered light. (b) Interpolated scattered light pattern.}
  \label{fig:scatter}
\end{figure*}

MINERVA employs a custom reduction pipeline, written in Python, following the diagram in Figure~\ref{fig:pipeline}\footnote{Available at: https://github.com/MinervaCollaboration/minerva-pipeline}.  This pipeline converts raw CCD exposures into calibrated, one-dimensional spectra with counts as a function of pixel position, $f(x)$.  The pipeline is broadly subdivided into two categories: calibration and extraction.

\subsection{Calibration}

The first step of calibration is bias subtraction.  To minimize the introduced noise, the median stacked bias frames are averaged across each column, and the difference from the overscan subtracted.  In practice, the overscan correction is much less than the read noise and has negligible impact on the final solution.  Although we collect nightly dark current frames, the level of our dark current is negligible, thus we omit dark current subtraction from the active pipeline.

Before flat fielding the CCD, we first normalize the stacked slit flats. This retains the blaze function in the extracted spectra, and prevents it from affecting the pixel to pixel corrections.  To model the blaze, we first fit for the throughput variation across the slit based on the approach described in \citet{bern2015a}.  We fit a fourth order polynomial to the (cross-dispersion) cross-section of each slit flat, as illustrated in Figure~\ref{fig:slit_flat}. The sharp edges are not well-fit by this low-order polynomial and are instead determined using the median value of $\pm3$ pixels in the dispersion direction. Any values below 50\% of the local mean are not fitted.  We then interpolate these smoothed profiles along the dispersion direction to estimate the blaze function and remove it from the flats.  Finally, we divide the bias-subtracted CCD by the normalized slit flats.


Although the slits block much of the scattered light, a small amount will typically reach the CCD.  We use the smoothed slit flats to mask the fiber bundles and view only the inter-fiber spaces where scattered light dominates (left of Figure~\ref{fig:scatter}).  
We interpolate first across each column, then across each row with a 6th order Legendre polynomial.  From this initial scattered light estimate, we mask cosmic rays and any other defects.  We then repeat our interpolation with 12th order Legendre polynomials to generate a final estimate, such as that shown on the right of Figure~\ref{fig:scatter}, and subtract this pattern from the exposure.  This fit is poorly constrained in the blue orders (top of the image) because of tight fiber packing, thus we constrain the interpolated image to be within $\pm50$ counts of the background median. We also include a variable background term during extraction (see section~\ref{sec:oe} for details) to account for any errors in scattered light subtraction.

These steps are summarized in this formula for the calibrated data,
\begin{equation}
  \text{D}_{\text{cal}} = \text{gain}\times\left((\text{D}_{\text{raw}} - \text{Bias} - \text{Scatter})/\text{Slit Flat}\right).
  \label{eqn:calibrate}
\end{equation}
Here, `$\text{D}$' is the science exposure, `Bias' is the stacked bias exposure, `Scatter' is the interpolated scattered light image, and `Slit Flat' is the stacked and smoothed slit flat.  We also multiply by the CCD gain of 1.3.  We leave the result in raw counts, rather than performing an absolute flux calibration which can be challenging and uncertain \citep{suzu2003a, hens2007a}.

In addition to the CCD calibration, we perform an initial wavelength calibration using a series of thorium-argon (ThAr) exposures taken on November 23, 2016, and an adaptation of the procedure described in \citet{murp2007a}.  We first extract each arc calibration frame, then visually compare the emission lines to the High Accuracy Radial velocity Planet Searcher (HARPS) atlas\footnote{http://www.tng.iac.es/instruments/harps/data/ThAr\_Atlas.pdf}.  Within each order, we determine the pixel position and wavelength of five well-spaced lines and use these as references.  We then run an algorithm to find $\lambda(x)$ for each order, using as many good lines as possible.  We define ``good'' lines as those which fulfill the following criteria:
\begin{enumerate}
    \item No known emission lines within 6 pixels of the line center.
    \item Peak counts are more than $4\,\sigma$ above the background noise level.
    \item Peak counts are at least 1\% below the CCD saturation level.
\end{enumerate}
We find the centroid of each good line using a Gaussian profile and fit the wavelengths, $\lambda$, of each centroid to a second order polynomial in pixel position, $x$. This initial wavelength solution, $\lambda(x)$, is passed to the Doppler pipeline of \citet{wils2019a}, providing a good starting estimate for the precise iodine wavelength calibration.

\subsection{Extraction (Optimal)}
\label{sec:oe}

\begin{figure}
    \centering
    \includegraphics[width=0.5\textwidth]{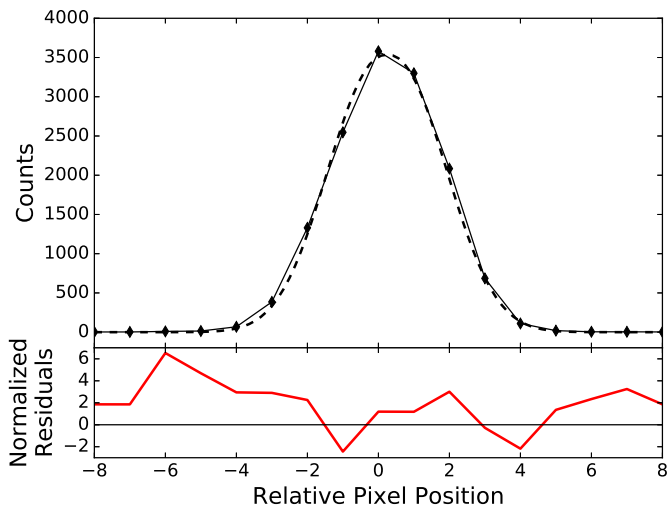}
    \caption{Profile fit of a fiber flat.  The top panel shows the counts as a function of pixel position with the model profile shown with a dotted line.  The bottom panel shows the normalized residuals, $(I(y)-p(y))/\sigma$.}
    \label{fig:oe_fit}
\end{figure}

Given calibrated CCD exposures, the spectrum is found by solving for counts as function of pixel position, $\bm{f}(\lambda(x))$.  This is found by solving the matrix equation
\begin{equation}
  \text{D}_{\text{cal}}(x,y) = \sum_{\lambda}\bm{P}_{\lambda}(x,y)\bm{f}(\lambda(x)) + \bm{N}(x,y).
  \label{eqn:extraction}
\end{equation}
Here, D$_{\text{cal}}(x,y)$ is the photon count as a function of the column, $x$, and row, $y$, on the calibrated CCD.  With MINERVA's alignment $x$ roughly corresponds to the dispersion direction and $y$ to the cross-dispersion direction.  The wavelength $\lambda(x)$ depends on the wavelength solution found during calibration.   The noise is given by $\bm{N}(x,y)$ and, for MINERVA, is generally dominated by Poisson photon counting statistics.  The instrument response, or point spread function, is given by the matrix $\bm{P}_{\lambda}(x,y)$.

Optimal extraction \citep{horn1986a}, simplifies equation~\ref{eqn:extraction} by assuming that intensity (photon count) is a separable function of $x$ and $y$ such that $\text{D}_{\text{cal}}(x,y) = \bm{I}(x)\bm{I}(y)$.  This allows us to express the observed cross-dispersion intensity as a function of the flux at each wavelength $\lambda(x)$ through
\begin{equation}
  \bm{I}(y) = \bm{p}(y,\lambda(x))f_{\lambda(x)} + \bm{n}(y)
  \label{eqn:oe}    
\end{equation}
Now the profile matrix $\bm{p}$ is a one-dimensional function of $y$ at each $\lambda(x)$.  We solve equation~\ref{eqn:oe} for the flux at each position, then join to find the spectrum $\bm{f}(\lambda(x))$.  Slight variants of this technique are applied in many EPRV reduction pipelines \citep{angl2012a, chak2014a, zech2014a, bern2015a, roy2016a, brah2017a, zech2018a, bech2019a, ma2019a}.

We empirically determine $\bm{p}$ from stacks of high signal-to-noise fiber flats collected on February 26th, 2017.  As with the ThAr exposures, these flats cannot be collected regularly due to the instrument configuration.  We have found, however, that the cross-dispersion profile of the KiwiSpec spectrograph does not evolve significantly on timescales of $\approx1\,\text{year}$ \citep{wils2019a} and so any bias due to instrument drift is small.  If, however, the profile shape evolves substantially in the future, this approach may require an update.



We explored several profile choices including a Gaussian, Moffat, B-spline interpolation, Lorentzian, and Voigt.  Of these, we found the best performance, judged by the $\chi^2$ statistic across the CCD, was achieved with a modified Gaussian profile
\begin{equation}
    \bm{p}(y) = I_0\, e^{\textstyle \left(-0.5\left(\frac{|y-y_c(x)|}{\sigma(x)}\right)^{p(x)}\right)}
    \label{eqn:mod_gauss}
\end{equation}
The centroids, $y_c(x)$, the width, $\sigma(x)$, and a generalized exponent, $p(x)$, are each fitted within each order as a smoothly varying B-spline
along the dispersion ($x$) direction.
MINERVA's circular fibers are nearly symmetric in the $x$ and $y$ directions and slightly broadened in the center with a characteristic exponent of $2.1 \la p(x) \la 2.3$.  A typical profile fit and residual are illustrated in Figure~\ref{fig:oe_fit}.  


The centroids $y_c(x)$ serve as estimates for the traces of each order.  For each science exposure, we re-fit the trace to account for drift over time.  
We use the fiber flat trace to flag and report anomalously large shifts.

With the profile in equation~\ref{eqn:mod_gauss}, we perform optimal extraction along each column.  Because some of the MINERVA orders, particularly in the blue side, are closely packed, there can be significant fiber-to-fiber cross talk.
To account for this, we extract an entire column at once by generalizing equation~\ref{eqn:oe} as a sum of fibers.  We also include a twelfth-order polynomial background across the column, which accounts for any errors in the scattered light subtraction algorithm.  We hold the trace centers and profile shape fixed, making this a linear fit.

The noise, $\bm{n}(y)$ is first estimated from the data, then updated to use the model noise as in \citep{horn1986a}.  We also mask cosmic rays using the critera
\begin{enumerate}
    \item The highest $\chi^2$ error pixel in the cross-sectional fit to each fiber.
    \item More than $5\sigma$ above the noise limit.  This equates to less than one false positive rejection per exposure.
\end{enumerate}
We iterate this rejection algorithm up to three times per column, stopping earlier if no changes are detected.  If a particular fiber cross-section has at least one cosmic ray rejection, we mask that column $x$ in $f(\lambda(x))$, excluding it from the RV analysis.

After extraction, we apply barycentric corrections based on FAU flux-weighted mean times.  Our pipeline also automatically detects and reports any low exposures or significant night-to-night deviations.
We save the extracted counts, ThAr wavelength solution, inverse variance, and cosmic ray mask for each pixel, order, and telescope.  The total extraction time is $\approx300\,\text{s}$ per exposure on the University of Utah Center for High Performance Computing\footnote{https://www.chpc.utah.edu/}.



\section{Implementing Spectro-Perfectionism}
\label{sec:sp}

We now move on to the challenge of extracting the spectra with spectro-perfectionism.  This amounts to solving equation~\ref{eqn:extraction} for the general case of a non-separable profile.  The full technique is described in \citet{bolt2010a}, but the authors conclude with the conundrum that SP extraction, as posed, cannot be solved numerically by present-day computers.  We describe adaptations that allow us to perform SP extraction.  SP extraction also requires knowledge of the two-dimensional point spread function (PSF) across the CCD, which is challenging to determine to high precision.

\subsection{Reducing Calculation Time}

\begin{figure*}[t]
\gridline{\fig{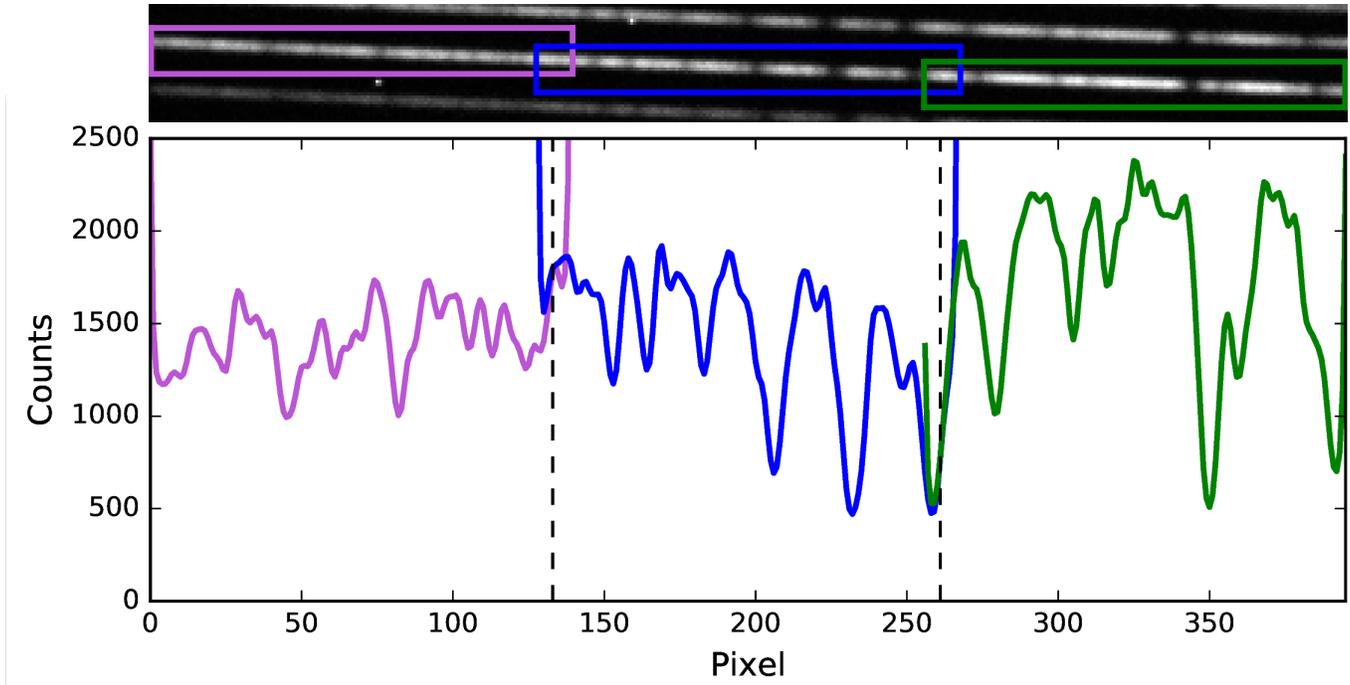}{1.0\textwidth}{}}
  \caption{Overlapping boxes drawn atop part of the calibrated image of one fiber. Extracted counts are shown below the fiber image.  The dotted line indicates where the spectra are joined.  We discard the invalid solutions at the edges of each box.}
  \label{fig:sp_boxes}
\end{figure*}

The least squares solution to equation~\ref{eqn:extraction} is given by
\begin{equation}
    \bm{f}_{\lambda} = (\bm{P}^T\bm{N}^{-1}\bm{P})^{-1}\bm{P}^T\bm{N}^{-1}\text{D}
    \label{eqn:sp_soln}
\end{equation}
where we drop the $x$, $y$, and $\lambda$ variables for clarity. \citet{bolt2010a} point out, however, that even for a modestly sized CCDs like MINERVA's, the matrix $\bm{P}$ is of size $\approx10^{12}$ which cannot be stored in memory, nor inverted.  We cannot simultaneously extract the entire CCD.

Several studies have explored practical implementations of spectro-perfectionism or other two-dimensional extraction algorithms \citep{zhu2011a, yu2014a, guan2015a}.  \citet{guan2015a} were able to extract an entire CCD exposure by solving for perturbations to an initial estimate and locally extracting blocks of the CCD. We present here a different method, adapted from work performed by the Dark Energy Spectroscopic Instrument (DESI) collaboration \citep[Stephen Bailey and Nicol\'{a}s Busca, private communication]{desi2016a}. Like \citet{guan2015a}, we perform localized extraction, but we do not require an initial estimate.  

Although the point spread function is, in principle, infinite in extent, it only contributes significant weight in a relatively small region.  For example, the counts in the wings are negligible beyond $\approx6\,\text{pixels}$ of the line centroid of the MINERVA PSF, as can be seen in Figure~\ref{fig:oe_fit}.  Thus we can perform a valid SP extraction on a small region of the CCD at a time.  This dramatically shrinks matrix sizes, enough that the solution is computationally tractable.


We divide each fiber into a series of boxes, as shown in Figure~\ref{fig:sp_boxes} (top).  We then extract the spectrum within each box.  The local PSF approximation is, however, invalid near the left and right edges of the boxes where adjacent points contribute significant flux.  Thus we overlap the boxes we use for extraction by 12 pixels and eliminate the outer $\pm6$ pixels of extracted flux from each box.  We then join the extracted counts from the inner regions of the boxes to find the spectrum for each fiber as illustrated in Figure~\ref{fig:sp_boxes} (bottom).  The accuracy is insensitive to the number and size of the boxes, so these may be chosen to optimize speed of calculation.  For MINERVA we use 16 boxes per order, each 140 pixels long

We also find that the profile matrix, $\bm{P}$, is mostly zero within each box.  Using Python's \texttt{sparse} package, we further improve speed with sparse matrix calculations.
This gives us a total extraction time of roughly $30\,\text{minutes}$ on an Intel Core i7-4700HQ CPU with 16Gb of memory.

We did not employ cosmic ray rejection during SP extraction, nor did we account for fiber-to-fiber cross-talk.  Cosmic rays have only a modest impact in MINERVA and we found evidence for only ${\sim}0.1\%$ fiber cross-talk.  SP extraction naturally minimizes the contributions from both of these effects because their shapes do not follow the PSF. The algorithm can easily be extended to include both effects, but this will generally increase the total computational time.



\subsection{Point Spread Function Fitting}

\begin{figure}[t]
\gridline{\fig{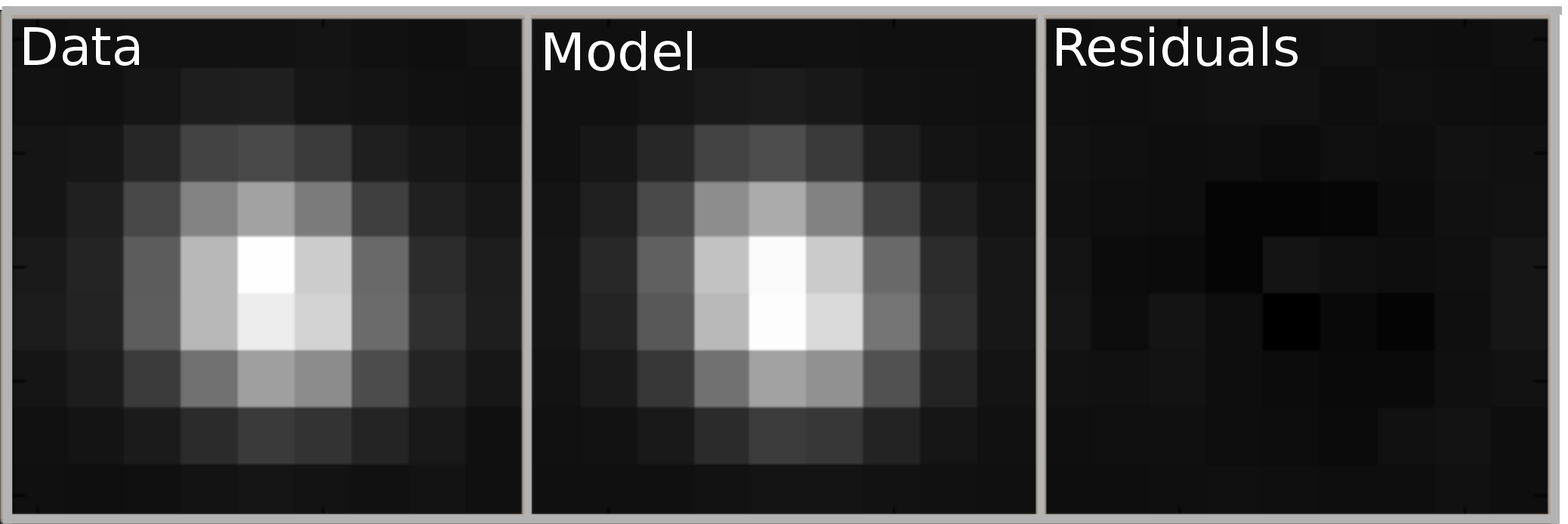}{0.45\textwidth}{(a)}}
\gridline{\fig{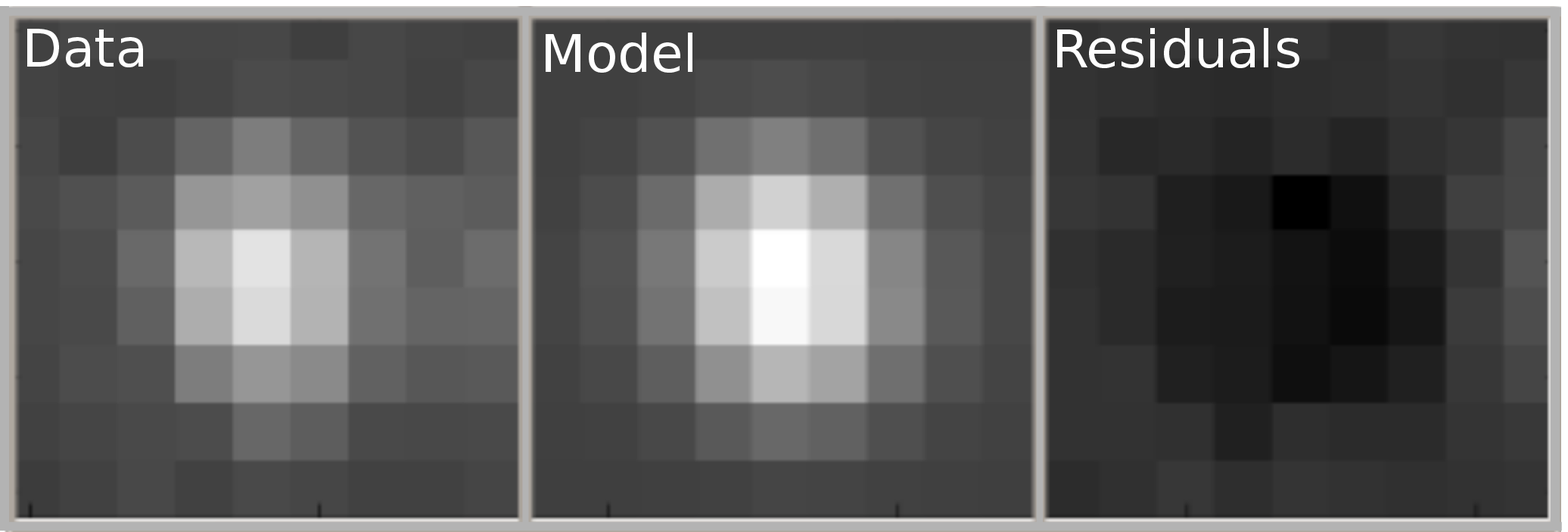}{0.45\textwidth}{(b)}}
  \caption{Point spread function (PSF) fits to thorium-argon emission lines using a two-dimensional Gauss-Hermite polynomial model (equation~\ref{eqn:gh}).  (a) Data, model, and residuals for a well-isolated line with excellent PSF fitting. (b)  Data, model, and residuals for a line with a neighbor contaminating the flux and distorting the fitted PSF.}
  \label{fig:thar_psf}
\end{figure}

As with finding $\bm{p}$ for one-dimensional optimal extraction, we must determine a profile $\bm{P}$ (the PSF of the CCD) for spectro-perfectionism.  However, in the two-dimensional case, the fiber flats are not suitable for the determination of the PSF because the profile is ill-constrained in the dispersion direction.  The easiest way to estimate the PSF is by fitting to the shape of unresolved emission lines.  For MINERVA, our best source of such lines is the ThAr arc lamp used for the initial wavelength calibration.

For our profile shape we adopt two-dimensional Gauss-Hermite polynomial, following the form proposed by \citet{pand2011a}.  We again modify the Gaussian to have a free exponent $p$ giving the form
\begin{multline}
G_{mn}(x,y) = \frac{1}{2\pi\sigma^p}H_m\left(\frac{x'}{\sigma}\right)H_n\left(\frac{y'}{\sigma}\right) \\ \times e^{\textstyle \left(-0.5\left(\frac{|x'|^{p} + |y'|^p}{\sigma^p}\right)\right)}.
\label{eqn:gh}
\end{multline}
For the width, $\sigma$, and power, $p$, we use the fitted values from the one-dimensional profile of equation~\ref{eqn:mod_gauss}.  The coordinates $x'$ and $y'$ are transformed from the pixel coordinates $x$ and $y$ by
\begin{align}
x' &= q(x-x_c)\cos{\phi} - (y-y_c)/q\sin{\phi} \\
y' &= q(x-x_c)\sin{\phi} + (y-y_c)/q\cos{\phi}
\label{eqn:xprime}
\end{align}
This allows rotated elliptical profiles with axis ratio $q$ and position angle $\phi$ at the centroids $x_c$ and $y_c$. The Hermite polynomials $H_m$ and $H_n$ are of orders $m$ and $n$ where we restrict to $m+n \leq 4$ and are given by  
\begin{equation}
H_n(x) = (-1)^ne^{\frac{x^2}{2}}\frac{d^n}{dx^n}e^{-\frac{x^2}{2}}.
\label{eqn:hermite}
\end{equation}
We experimented with including a Lorentzian wing component \citep{pand2011a}, but the fit improvement was not significant enough to justify the added complexity.  We also tested radial B-spline models with an elliptical component which performed well, but we achieved our best performance, as judged by the $\chi^2$ metric, with the Gauss-Hermite model.

Like the one-dimensional cross-section, the PSF shape changes slowly across orders.  Thus we impose that all the parameters, including the Gauss-Hermite coefficients, must be fitted by a second order polynomial along each order.  Only the Hermite coefficients can be found linearly, so we adopt an iterative procedure and solve for the PSF for each by taking the following steps:

\begin{enumerate}
\item Normalize each ThAr emission line, centered within a $13\times 13$ pixel box;
\item Fit for the Gaussian terms $x_c$, $y_c$, $\sigma$, $p$, $q$, and $PA$ using the Python package \texttt{lmfit} \citep{newv2014a};
\item Linearly solve for Hermite coefficients;
\item Iterate steps 1--4 until subsequent fits have a change in reduced $\chi^2$ of $\Delta\chi^2_r < 0.001$.
\end{enumerate}
Typical best-fit models are shown in Figure~\ref{fig:thar_psf}.

This procedure was time intensive, requiring roughly 1 hour of cpu time per fiber on an Intel Core i7-4700HQ CPU with 16Gb of memory.  For MINERVA, however, we only have one set of ThAr exposures, so repeated fitting was neither necessary nor possible.  We use the fitted PSF from the ThAr exposures on November 23, 2016 for every SP extraction.  Because the MINERVA PSF is stable on a timescale one year \citep{wils2019a}, the static PSF provides a sufficiently accurate solution.  However, to account for drift on timescales longer than a year, more frequent ThAr exposures would be required.

There are, however, limitations of ThAr exposures.  Intensities vary widely between lines, spacing is inconsistent between orders, and many lines are blended. As with wavelength solution fitting, we filtered for good lines, adopting a slightly stricter criterion of lines with no strong neighbors within $9$ pixels of the centroid.  We also cut out any lines with a poor $\chi^2$ fit to a Gaussian, which helps eliminate blended lines.  Nevertheless, some line blending still contaminated our fits, as seen in the bottom panel of Figure~\ref{fig:thar_psf}.

Furthermore, these stringent cuts severely limited the sampling within each order.  We retained $12 \text{--} 35$ lines per order to fit, which only provided coverage of $6-19\%$ of each 2048 pixel order.  Irregular line spacing within each order was also problematic, leaving some unconstrained regions several hundred pixels long.  These factors ultimately limited the precision of our empirical PSF.

\subsection{Re-convolution Matrix}
\label{sec:reconvolution}

One other subtlety of spectro-perfectionism, described in \citet{bolt2010a}, is the covariance between extracted points due to PSF blending.  In all but extremely high signal to noise cases, the extracted spectrum $\bm{f}(\lambda(x))$ exhibits extreme ringing, rendering the solution meaningless for assessing radial velocities.  This is overcome by the application of a ``re-convolution'' matrix $\bm{R}$ that smooths the extracted spectrum to give
\begin{equation}
    \tilde{\bm{f}}({\lambda}(x)) = \bm{R}\bm{f}({\lambda}(x))
    \label{eqn:sp_ftilde}
\end{equation}

The matrix $\bm{R}$ is not unique, and indeed a broad range of choices will return a smooth $\tilde{\bm{f}}({\lambda}(x))$. Although $\bm{R}$ is not unique, \citet{bolt2010a} supply a formulation to determine a matrix $\bm{R}$ that diagonalizes the covariance matrix.  This gives an extracted spectrum that is effectively broadened to the native resolution of the instrument. We use this method to construct $\bm{R}$ for MINERVA as part of the extraction procedure.

The re-convolution matrix plays an important role in the subsequent Doppler pipeline.  Doppler pipelines require a determination of the instrumental profile \citep[IP; e.g.,][]{wils2019a}.  The IP characterizes the broadening of the instrument in the dispersion direction.  For SP extraction, however, this broadening is completely characterized by $\bm{R}$.  Thus SP extraction brings an additional benefit of determining the IP during extraction.  Our pipeline calculates $\bm{R}$ for each box in Figure~\ref{fig:sp_boxes}, then averages within the box to estimate the corresponding one-dimensional IP.  The caveat is that for a truly unbiased answer, subsequent Doppler pipelines must use $\bm{R}$, not a conventionally determined IP.

\section{Spectro-Perfectionism Performance}
\label{sec:results}

\begin{figure*}[t]
\gridline{\fig{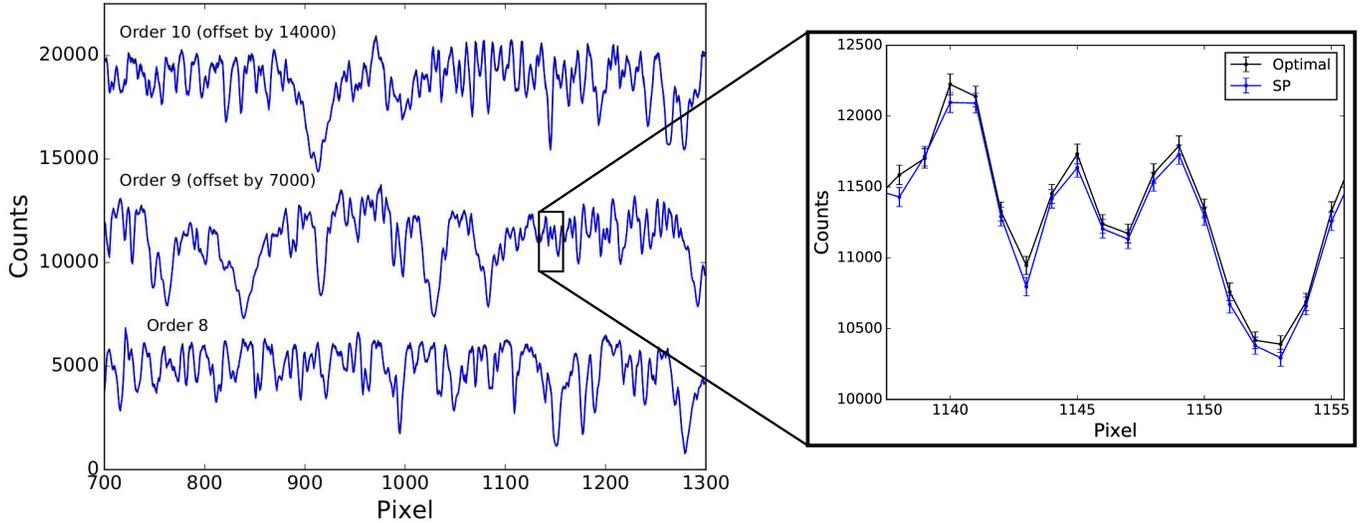}{1.0\textwidth}{(a)}}
\gridline{\fig{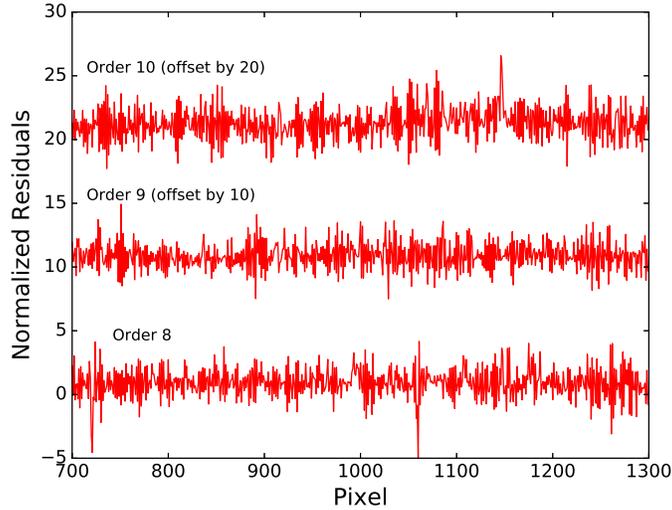}{0.5\textwidth}{(b)}}
  \caption{(a) Extracted spectra of HD122064 showing portions of three orders shown with a vertical offset for clarity.  The zoom in portion shows the spectra from optimal extraction (black) and spectro-perfectionism (blue).  (b) The error-scaled residuals for the same portions of spectra, showing the spectra are nearly statistically indistinguishable.}
  \label{fig:spectra}
\end{figure*}

\begin{figure*}[t]
\gridline{\fig{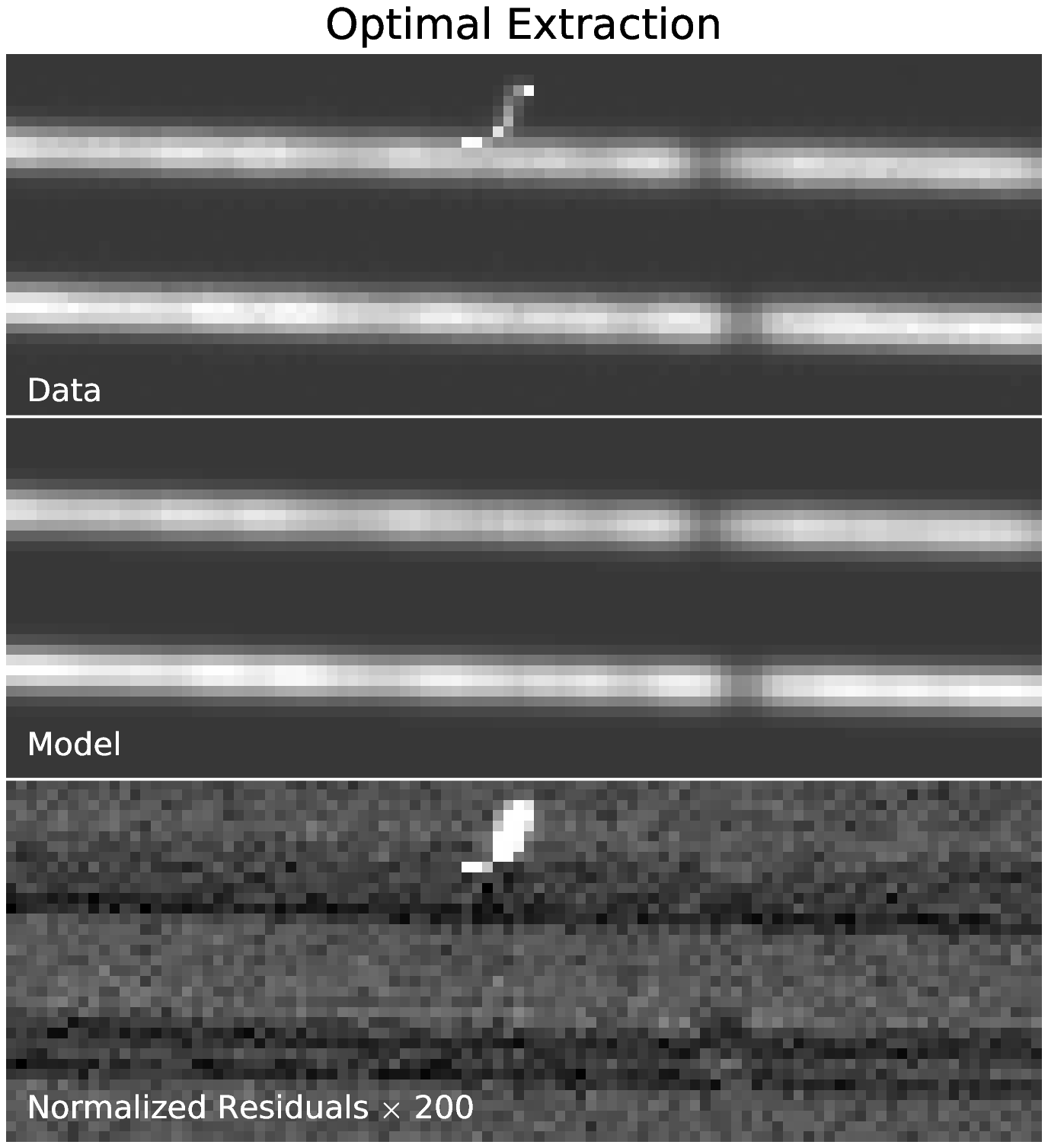}{0.5\textwidth}{(a)}
          \fig{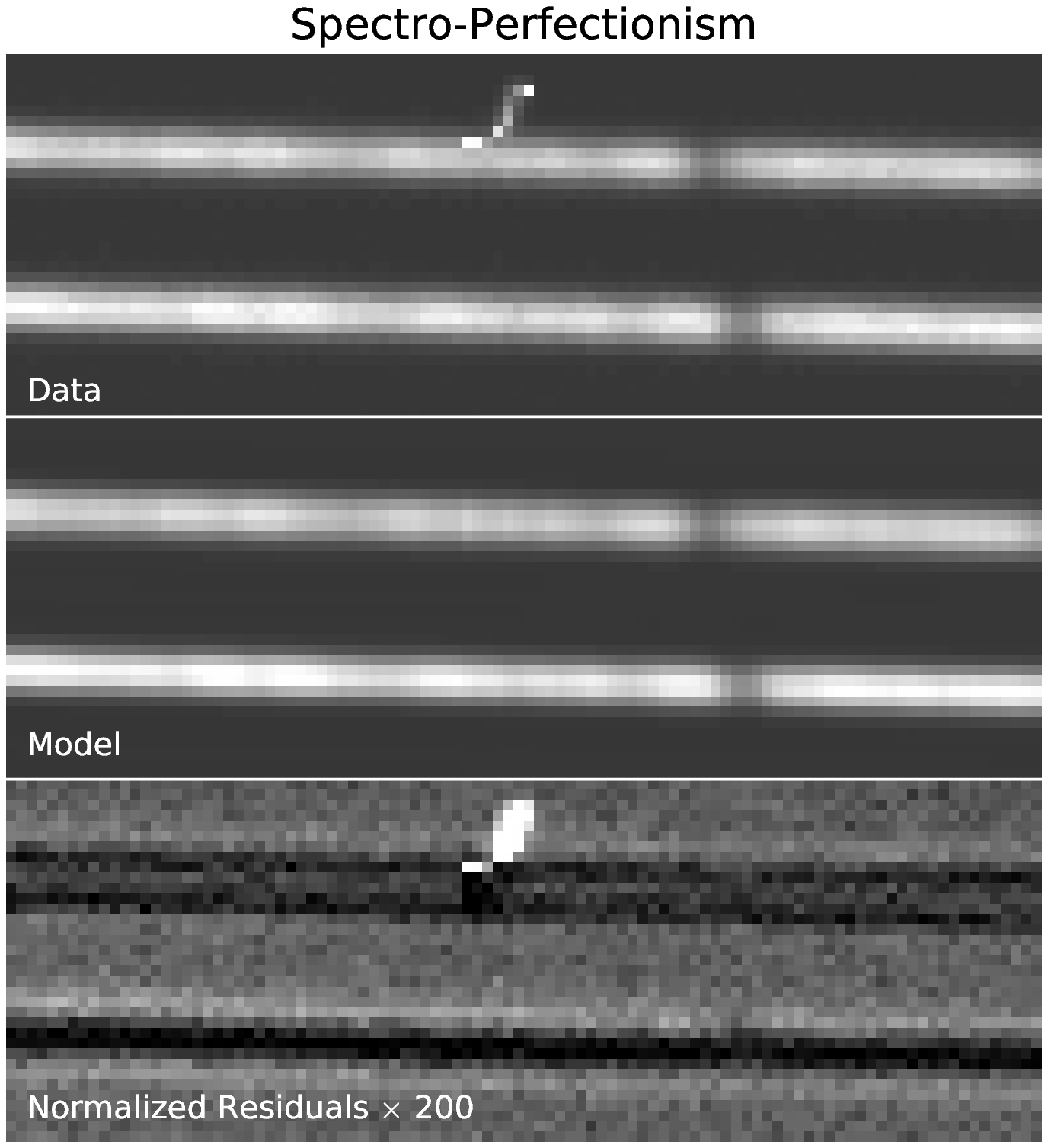}{0.5\textwidth}{(b)}}
  \caption{(a): Data (top), model (middle), and error-normalized residuals (bottom) for a CCD model generated during optimal extraction.  (b)  Data (top), model (middle), and error-normalized residuals (bottom) for a CCD model generated during spectro-perfectionism extration.  The vertical scales are the same showing that SP extraction has more structure in the residuals.  The cosmic ray hit near the center of the top fiber has a negligible impact on adaject pixels, even in SP extraction without explicit cosmic ray masking.}
  \label{fig:ccd_model}
\end{figure*}

There is no single metric to evaluate the relative performance of optimal extraction and spectro-perfectionism. We can, however, aggregate several outputs to compare their accuracy and precision.



\subsection{Spectral Comparison}
\label{sec:spec_compare}


We directly compare the extracted spectra from HD122064 taken on May 20th, 2016 in Figure~\ref{fig:spectra}. By eye the two spectra are nearly indistinguishable.  We quantify the statistical differences through the metric of the normalized residuals
\begin{equation}
    N = \frac{2(\bm{f}_{\text{oe}} - \bm{f}_{\text{sp}})}{\sqrt{\sigma^2_{\text{oe}}+\sigma^2_{\text{sp}}}}.
    \label{eqn:oe_sp_noise}
\end{equation}
Here, $\bm{f}_{\text{oe}}$ and $\bm{f}_{\text{sp}}$ are the extracted counts of each, and are divided by the average error estimate.  The normalized residuals are shown in Figure~\ref{fig:spectra} (bottom).  For equivalent spectra with Gaussian noise, the standard deviation of $N$ should approach 1.  Over the entire CCD, we found $\sigma_{N}\approx1.5$, when rejecting outliers with $|N|\geq5.0$.  We also found a slight count offset between the two spectra of $(f_{oe}-f_{sp})/f_{oe} = 0.014$, which may be due to differences in background modeling.

This indicates that the extracted spectra are not identical.  This fact alone, however, does not tell us which spectrum is a better representation of the ``true'' spectrum.  Furthermore, we do not expect the spectra to be in perfect agreement as a result of the re-convolution matrix (see section~\ref{sec:reconvolution}).  The SP spectrum is broadened by our re-convolution matrix while the optimal extraction carries an unknown signature of the natural instrument broadening.  These effects are similar, but not identical.
Nevertheless, the spectra display remarkable consistency which assures us that our SP extracted spectrum is reliable.


\subsection{CCD Models}
\label{sec:ccd_compare}

We also evaluate the accuracy with which each method models the calibrated CCD.  From the profile matrices and fitted flux, we can generate a model of the CCD through
\begin{equation}
    \bm{M}(x,y) = \sum_{\lambda,\text{fib.}}\bm{P}(x,y,\lambda,\text{fib.})\bm{f}(\lambda,\text{fib.})
    \label{eqn:ccd_model}
\end{equation}
where $\bm{M}$ is the model CCD, $\bm{P}$ is our PSF as a function of wavelength and fiber (fib.), and $\bm{f}$ is our extracted flux (without re-convolution).  Equation~\ref{eqn:ccd_model} holds for spectro-perfectionism only, but a similar model can be constructed for optimal extraction.

We find the $\chi^2$ per degree of freedom for each method by comparing $M$ to the calibrated CCD data $\text{D}_{\text{cal}}$.  
This value varies between exposures, but for a typical exposure of HD122064 we found $\chi^2_{r,\text{oe}}=5.8$ for optimal extraction.  The value for spectro-perfectionism is $\chi^2_{r,{\text{sp}}}=14.2$, roughly a factor of 2.5 higher across the CCD.
We can quickly visualize this difference by comparing the data, model, and normalized residuals for a typical portion of the CCD in Figure~\ref{fig:ccd_model}.  Although both extraction methods have some minor residual structure, there is more evident in the SP images and in particular a stronger mismatch between trace centers and wings.  This indicates our PSF model does not completely capture the true shape of the projected light profile.

\subsection{Radial Velocity Precision}

\begin{figure*}[t]
\gridline{\fig{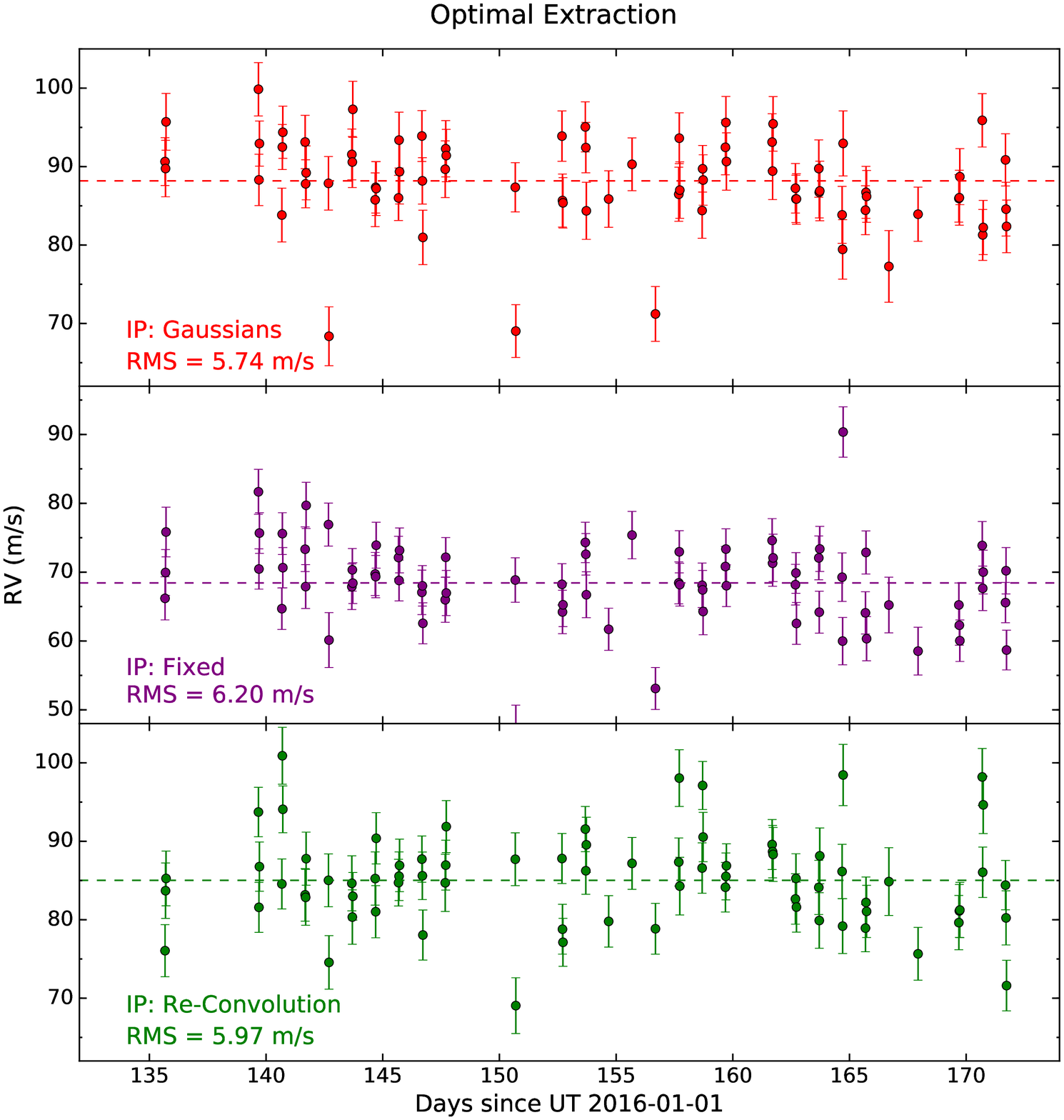}{0.5\textwidth}{(a)}
          \fig{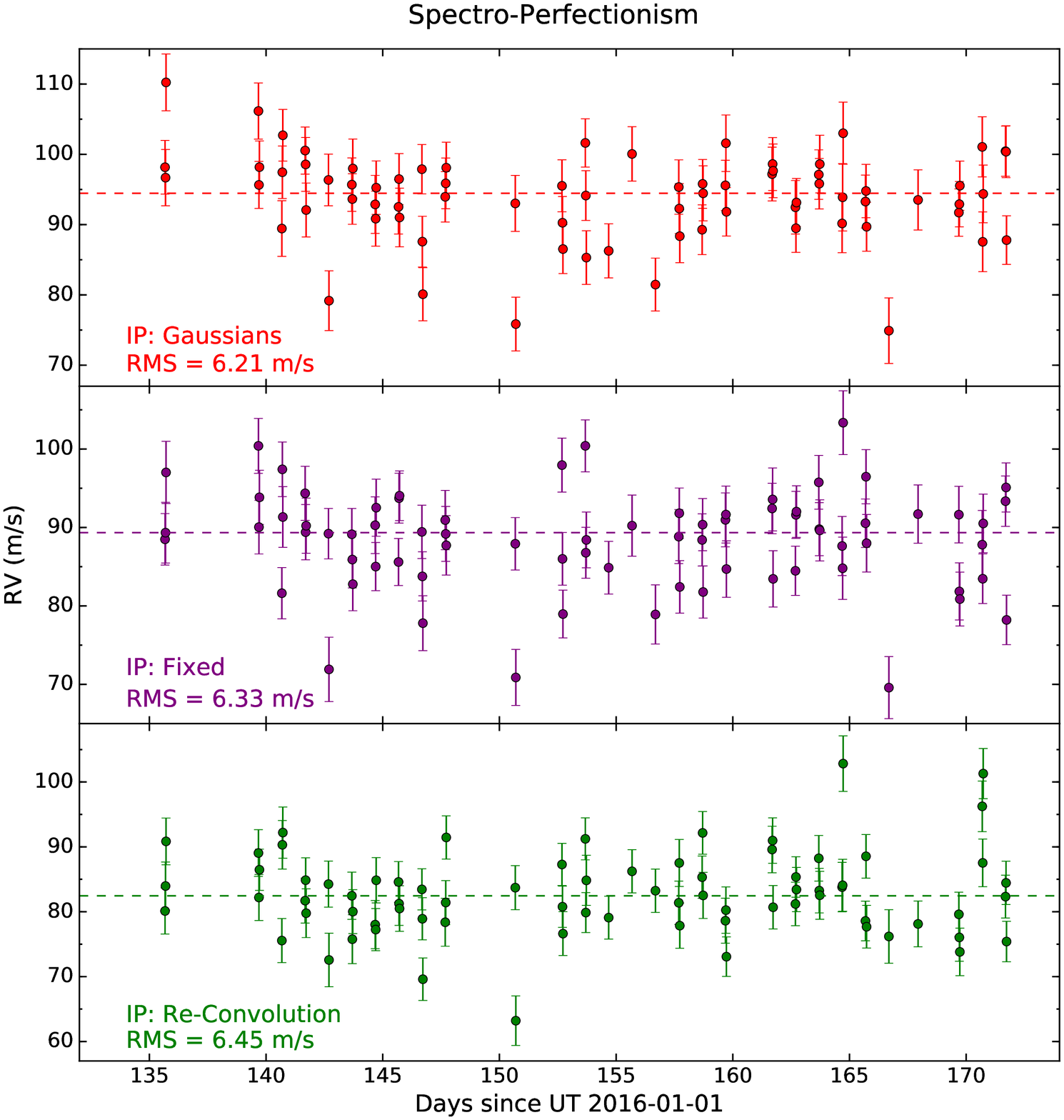}{0.5\textwidth}{(b)}}
  \caption{(a) Radial velocities determined for HD122064 on telescope 2 of MINERVA using the optimal extraction spectra.  (b)  Radial velocities determined for HD122064 for telescope 2 using the spectro-perfectionism spectra.  The median RV is indicated with a dashed line.  For both extraction methods, radial velocities are calculated using three different instrumental profiles: the sum of Gaussians (top, red), the fixed spline (middle, purple), and the re-convolution matrix $\bm{R}$ (bottom, green).}
  \label{fig:rvs}
\end{figure*}

\begin{deluxetable}{l|cc}
\tablecaption{Radial velocity precision for optimal extraction and spectro-perfectionism for a variety of instrument profiles (IPs) including the re-convolution matrix $\bm{R}$.  These results use the signal from telescope 2 only, binned by each 6 chronological exposures.\label{tab:rv_precision}}
\tablehead{
\colhead{IP} & \colhead{OE precision (m/s)} & \colhead{SP precision (m/s)}
}
\startdata
Gaussians & 2.25 & 2.31 \\
Fixed & 2.32 & 2.32 \\
$\bm{R}$ & 2.57 & 2.58 \\
\enddata
\end{deluxetable}

Finally, the most important metric is the relative radial velocity precision of the methods.
For our radial velocity test standard we selected the star HD122064, an inactive type K4V star with magnitude $V=6.5$.  We use a total of 119 exposures taken in May and June 2016, which we reduce with our pipeline, using both optimal extraction and spectro-perfectionism.
We run each extracted spectrum through the Doppler pipeline described in \citet{wils2019a}.  This pipeline finds a precise wavelength solution from the iodine absorption spectrum, forward models a stellar template convolved with the instrument profile, and generates RV values and errors separately for each telescope within an exposure.  

In Figure~\ref{fig:rvs}, we show RVs for each exposure calculated under a range of instrument profiles (IPs) for both the optimal extraction pipeline and the spectro-perfectionism pipeline.  In our sample, telescope 2 (T2) had the best throughput and so we restrict our reported results to T2.  Following the procedure described in \citet{wils2019a}, we calculate the Allan variance of the RVs, binned by each six consecutive good exposures, giving the precision measurements shown in Table~\ref{tab:rv_precision}. We selected a binning of six because systematic errors influence the RV precision at larger bin sizes. For each extraction method, we include results for three different instrument profiles including two from \citet{wils2019a}: the sum of Gaussians and a fixed profile from B-splines of the daytime sky.  We also use the re-convolution matrix $\bm{R}$ calculated during spectro-perfectionism extraction.

We find that for the re-convolution matrix and fixed IPs the precision is nearly identical between optimal extraction and spectro-perfectionism.  Both extraction methods show the best precision for the sum-of-Gaussians IP, and, using this IP, optimal extraction precision exceeds spectro-perfectionism precision.  Optimal extraction also shows the lowest precision floor, under further binning, for all three IPs. The relative performance between the two pipelines varies little with bin size, although in some cases spectro-perfectionism is favored. In general the calculated precision values are remarkably similar.



\section{Discussion and Conclusions}
\label{sec:discussion}

We have developed a full raw reduction pipeline for the MINERVA project. Although our standard pipeline uses the optimal extraction algorithm, we have developed the option for a computationally tractable method of spectro-perfectionism.
SP extraction is still computationally intensive, requiring approximately 30 minutes to extract a $2052\times2048$ pixel CCD with 112 fibers on an Intel Core i7-4700HQ CPU with 16 Gb memory.  This is slow compared to the $\approx1\,\text{minute}$ timescale achieved for fast optimal extraction \citep{ritt2014a}.  An hour or less per exposure is, however, a modest task for a high performance cluster, making SP extraction a realistic option for EPRV pipelines.  Furthermore, we have not optimized for the speed of SP extraction, so performance gains can be expected.  The extraction times for SP will, however, increase for larger detectors or broader PSFs.

We found comparable performance between the optimal extraction and spectro-perfectionism pipelines. We did, however, find a slightly better radial velocity precision with spectra from optimal extraction.  Additional work is needed before SP extraction will be favored for use in the MINERVA pipeline.
Nevertheless, the consistency of our findings in Table~\ref{tab:rv_precision}, show that spectro-perfectionism can achieve comparable performance to optimal extraction at the level of ${\sim} 2.3\,\text{m/s}$ precision.

For MINERVA, neither optimal extraction or SP extraction yet shows an RV precision at the instrument's capacity of $0.8\,\text{m/s}$.  \citet{wils2019a} suggest that a stellar template mismatch and instrumental profile imprecision may be the dominant limitations to MINERVA's precision.
As improvements are made to the RV pipeline, the relative performance between optimal extraction and SP extraction may change.
Further investigation is warranted, not solely for MINERVA, but for any EPRV instrument with $<\,1\text{m/s}$ precision where SP extraction may yet achieve better performance than optimal extraction. 

We also find that the re-convolution matrix $\bm{R}$ provides the worst precision for SP extraction.  In principle, however, $\bm{R}$ should give the best SP performance.  This suggests that PSF fitting bias is playing a limiting role in the ultimate precision of spectro-perfectionism.
This is not surprising given the residuals in Figure~\ref{fig:ccd_model}, which indicate a mismatch between our empirical PSF and the true instrument PSF.  Based on these findings, accurate PSF fitting appears to be the primary challenge of spectro-perfectionism.  Improvements in PSF fitting may translate to superior RV performance for SP extraction.

PSF fitting is easiest with access to clean calibration frames, particularly those with evenly spaced emission lines across each order and little to no overlap between lines.  Fortunately, such images are obtained for precise wavelength calibration on many forthcoming instruments using either laser frequency combs \citep{murp2007b, li2008a, wilk2010a, jurg2016a, schw2016a} or unresolved Fabry-Perot etalons \citep{halv2014a, rein2014a, stur2017a, cers2019a}.  Thus SP extraction is naturally enabled for upcoming EPRV exoplanet surveys.  Fitting will, however, be further complicated on instruments with unusual fiber shapes such as octagons, rectangles, or D-shapes which are used for fiber scrambling on various instruments \citep[e.g.,][]{chak2014a, locu2015a, jurg2016a, stur2016a}. SP extraction is also possible on slit spectrographs, but nightly or sub-nightly profile fitting would likely be required due to a time-variable PSF.

Optimal PSF accuracy may require special exposures in addition to the wavelength calibration frames.  Such PSF calibration frames would ideally include widely separated emission lines shifted through a sequence sub-pixel steps.  In practice, some line overlap or larger step sizes between exposures may be acceptable, but this will require further investigation to determine.  

PSF modeling is also an important ingredient for ``extreme'' forward modeling in which RV precision is determined by modeling directly to the CCD.  Extreme forward modeling would be computationally challenging and technically difficult, but could maximize precision by circumventing extraction entirely \citep{bede2019a}.

For now, SP extraction, though challenging, is computationally tractable and shows comparable RV precision performance to optimal extraction.  This technique may be a valuable component in forthcoming EPRV pipelines for exoplanet detection and characterization. Spectro-perfectionism also has potential for extracting faint-object spectra of high-redshift galaxies, as originally envisioned in \citet{bolt2010a}, but excellent PSF calibration, such as that achieved in \citet{kos2018a}, would be crucial.

\acknowledgements
MINERVA is a collaboration among the Harvard-Smithsonian Center for Astrophysics, The Pennsylvania State University, the University of Montana, and the University of Southern Queensland. MINERVA is made possible by generous contributions from its collaborating institutions and Mt. Cuba Astronomical Foundation, The David \& Lucile Packard Foundation, National Aeronautics and Space Administration (EPSCOR grant NNX13AM97A), The Australian Research Council (LIEF grant LE140100050), and the National Science Foundation (grants 1516242 and 1608203). Any opinions, findings, and conclusions or recommendations expressed are those of the author and do not necessarily reflect the views of the National Science Foundation.

Funding for MINERVA data-analysis software development is provided through a subaward under NASA award MT-13-EPSCoR-0011.

M.A.C. acknowledges support from the National Science Foundation under grant AST-1614018.

The support and resources from the Center for High Performance Computing at the University of Utah are gratefully acknowledged

\software{Anaconda, \texttt{lmfit} \citep{newv2014a}}

\clearpage

\bibliography{bibtex_archive_master}
\bibliographystyle{aasjournal}

\end{document}